\documentclass[preprint,showpacs,showkeys,preprintnumbers,amsmath,amssymb]{revtex4}

\usepackage{graphicx}

\usepackage{dcolumn}
\usepackage{bm}

\begin{document}


\title{Temperature distribution in large Bi2212 mesas}

\author{A. Yurgens}
 \affiliation{Department of Microtechnology and Nanoscience (MC2), Chalmers University of Technology, SE-412 96 G\"oteborg, Sweden}

\date{\today}

\begin{abstract}
I numerically analyze Joule heating in large Bi2212 mesas while taking into account typical thermal conductivities and their temperature dependencies of all the materials involved in the heat dissipation and its removal. Such mesas are used in experiments on THz-range radiation.  The analysis shows that the temperature increases with bias current and is distributed unevenly along the mesas. The temperature of the mesa's middle part can even exceed $T_c$ at sufficiently high bias.  The non-uniform temperature distribution can possibly be important for synchronization of emission from different junctions in the mesas. The overall current-voltage characteristics are also calculated self-consistently showing a negative differential conductance in a wide range of currents.
\end{abstract}

\pacs{74.25.Sv, 74.50.+r, 74.72.Hs}
\keywords{Intrinsic Josephson junctions, THz radiation}
\maketitle

\section{\label{sec:Intro}Introduction}

The ac Josephson effect provides the basis for superconducting high-frequency detectors and tunable radiation sources with a voltage-to-frequency conversion parameter involving only fundamental constants. Josephson junctions made of high-temperature superconductors (HTS) allow for such devices operating in a wide temperature range $\sim 4 - 90$~K and high frequencies including the THz band.  There are not many continuous-wave tunable sources available at the low end of this band: $0.3-1$~THz (see a review article, Ref.~\cite{intro:THz_review}). The HTS Josephson junctions would be almost ideal radiation sources at these frequencies if they only had sufficiently high output power. Thin films of $\mathrm{YBa_2Cu_3O_{7+\delta}}$ (YBCO) have been the material of choice for making HTS Josephson junctions.  The Fiske- and external-cavity resonances were clearly seen in such junctions that witnessed for the presence of Josephson radiation~\cite{intro:Dag,intro:Edstam1}. However, the radiation from the individual junctions is weak, usually a few tens of pW, which limits their utility for practical applications.

Synchronization (or phase-locking) of many Josephson junctions is needed to boost the radiation power. A few YBCO Josephson junctions made on the same chip were demonstrated to phase-lock and thereby make the emission stronger~\cite{intro:Edstam2,intro:Iguchi2}. The synchronization looks easier to achieve for the so called intrinsic Josephson junctions (IJJ's) which are densely packed inside single crystals of another popular HTS, $\mathrm{Bi_2Sr_2CaCu_2O_{8+\delta}}$ (Bi2212)~\cite{intro:Kleiner}. The high-frequency radiation from these junctions was seen in several works~\cite{intro:Iguchi_IJJ,intro:Batov_IJJ}. However, the radiation was not strong either which indicated an absence of junctions phase-locking.

A relatively intense THz-range emission has only recently been detected from large-area and high mesas etched into Bi2212 single crystals~\cite{intro:THz_radiation}. A number of models have been proposed to explain the phase locking of IJJ's in these mesas ~\cite{intro:THz_model1,intro:THz_model2,intro:THz_model3}.  Yet stronger radiation has been measured~\cite{intro:THz_strong} at a current much higher than that in the original work of  Ref.~\cite{intro:THz_radiation}.  This current corresponds to a back-bending region of the overall current-voltage (I-V) characteristics, i.e. the region with a negative differential conductance. Joule heating is apparently high at this bias and therefore requires analysis of the heat balance and cooling efficiency.

In this work, I calculate the temperature distribution in large-area mesas.  The calculations include realistic thermal conductivities of all the materials involved in the heat generation and removal, as well as their temperature dependencies and anisotropy.   In other words, I numerically solve the non-linear diffusion equation~\cite{comsol}:
\begin{equation}
\nabla \left( { - k(T)\nabla T} \right) = \rho(T)j^2
\end{equation}
\noindent for mesas of typical geometry. Here, $k(T)$, $j$, and $\rho$ are the anisotropic thermal conductivity,the current density, and the anisotropic resistivity, respectively.

The calculations show that the temperature distribution in the large-area mesas is non-uniform for a wide range of currents. The non-uniformity increases with current.  At sufficiently high bias, the temperature in the center even exceeds the superconducting critical one, $T_c$, while the rest of the mesa is still superconducting.  The appearance of the normal-state regions can be important for synchronization of radiation from many IJJ's at high bias~\cite{intro:THz_strong}.

\section{\label{sec:geometry}Geometry of model mesa}

\begin{figure}
\includegraphics[width=8.5cm]{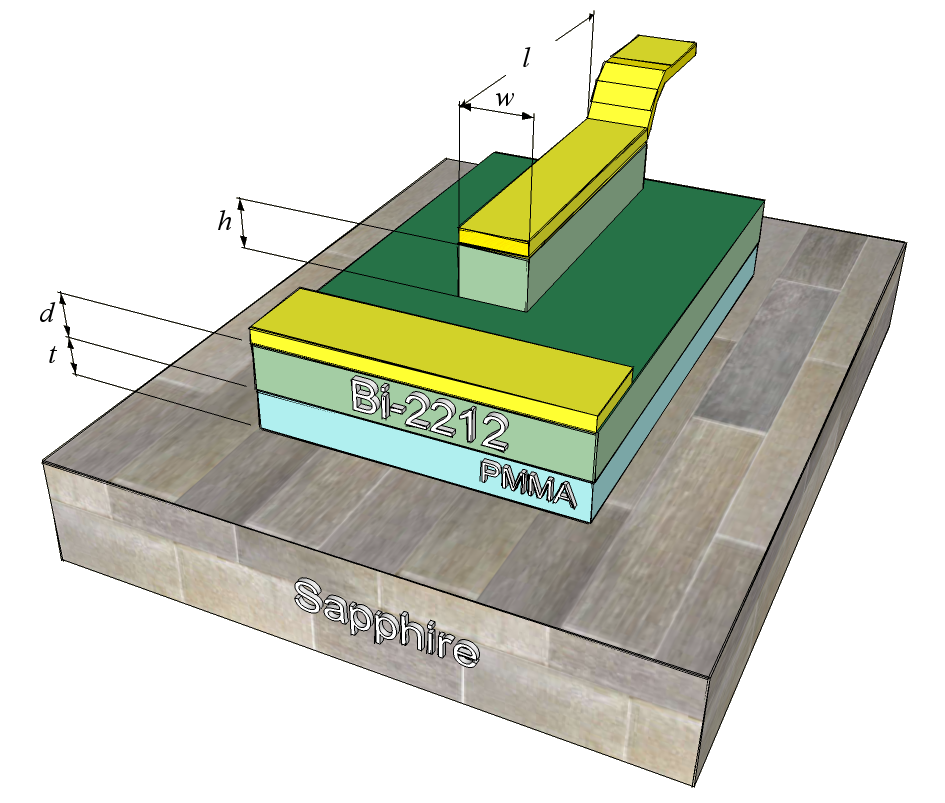}
\caption{\label{fig:geometry}(Color online) Schematic view of a mesa used in the calculations. The height $h=1$ or 2 $\mu$m; the width $w=50$ or $100$ $\mu$m. The length $l=300$ $\mu$m, the single-crystal thickness $d$ and the glue thickness $t$ are both 20 $\mu$m. Not in scale.}
\end{figure}

Fig.~\ref{fig:geometry} schematically shows a mesa on top of a Bi2212 single crystal. The crystal is taken to be $1\times 1$ mm$^2$ large and either 20 or 40 $\mu$m thick. The single crystal is attached to a sapphire substrate with a 20-$\mu$m-thick polymethylmethacrylate (PMMA) glue layer.
Due to the very high thermal conductivity of sapphire, the substrate surface is assumed to be at the bath temperature $T_0$. There is no cooling gas around the sample meaning that the heat dissipated in the mesa can only escape through the underlying single crystal and the PMMA layer.  The calculations are made for one and the same length of the mesa $l=300\ \mu$m. The mesa width $w$ (the height $h$) is either 50 or 100 $\mu$m (1 or 2 $\mu$m).  A 100-nm gold thin film is deposited on top of the mesa and the front part of the single crystal making up the bias electrodes (see Fig.~\ref{fig:geometry}). There is no interim resistive layer between the electrodes and Bi2212. All the simulations have been made using the finite-element software package COMSOL~\cite{comsol}.

\section{\label{sec:materials}Materials parameters}

For calculations, I use common materials parameters found elsewhere.  The anisotropic thermal conductivity $k(T)$ of Bi2212 single crystals was measured in a number of works, predominantly along the $ab$-plane~\cite{materials:k_ab_we,materials:k_ab_review,materials:k_ab_Ong,materials:k_ab_Kapitulnik}. To the best of my knowledge, there exist only two measurements of the thermal conductivity in the $c$-axis direction~\cite{materials:k_c,materials:k_c2}. It was found that the ratio $k_{ab}/k_c \sim 6$ in a wide temperature range and increases to 8 at low temperatures~\cite{materials:k_c}. In the calculations, I use $k_{ab}$ for pure Bi2212 adopted from Ref.~\cite{materials:k_ab_Kapitulnik} and simply assume that $k_c = k_{ab}/10$.

The $ab$-plane resistivity of Bi2212 is taken to be linear in temperature:
$
\rho _{ab}  = \theta (T-T_c,\delta T)T/75 \;\ [{\rm{\mu \Omega }}\,{\rm{m]}},
$
where $\theta (x,\delta x)$ is the Heaviside step function smeared over interval $\delta x$ and $T_c=86$ K is the superconducting critical temperature of the optimally doped Bi2212.  The Heaviside function is used to mimic the transition to the normal-metal state when the local temperature exceeds $T_c$.  The current-induced transitions within individual CuO planes~\cite{materials:Ic_ab} are ignored in these calculations.

\begin{figure}
\includegraphics[width=8.5cm]{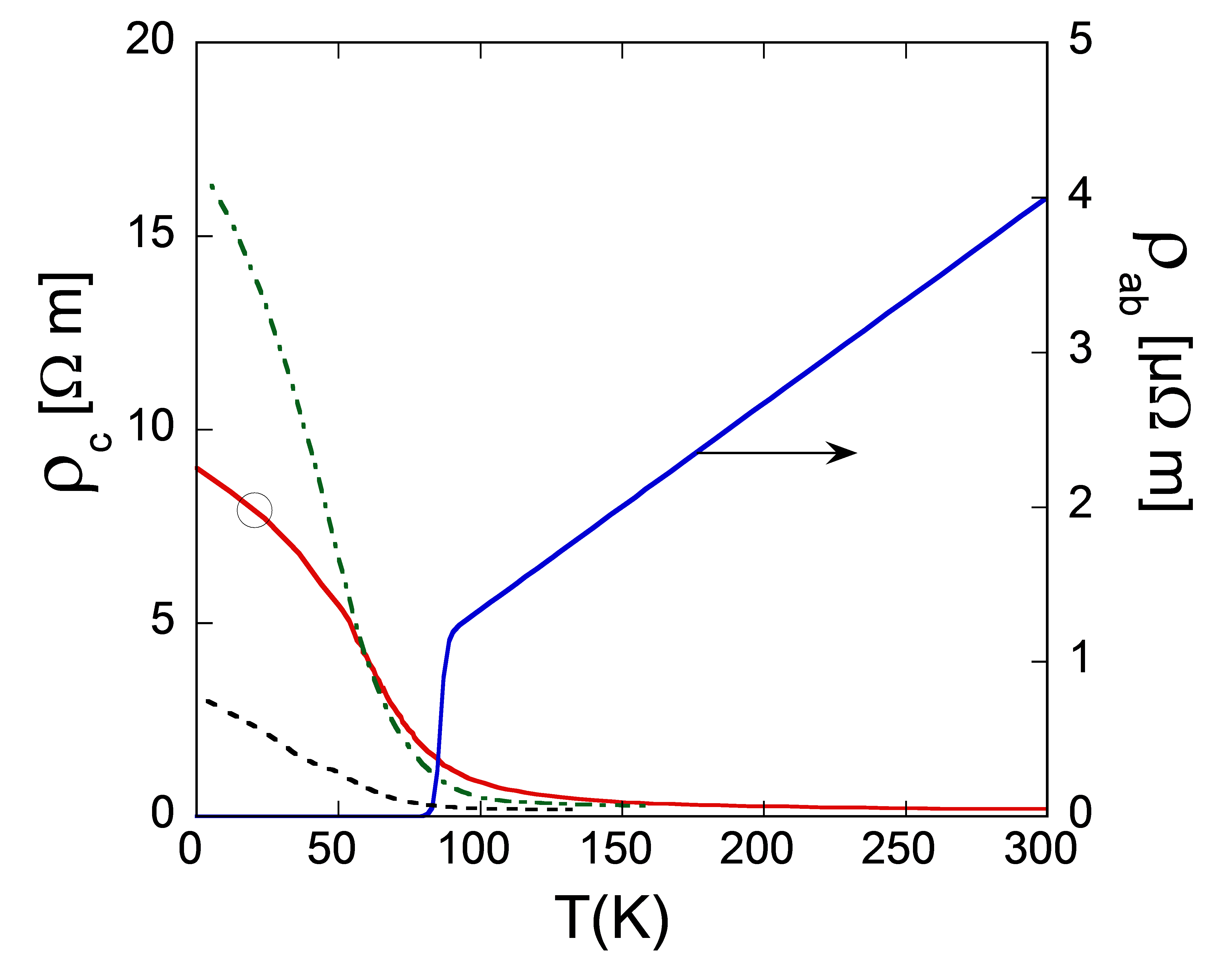}
\caption{\label{fig:Bi2212_param}(Color online) The $ab$- and $c$-axis resistivities assumed in the calculations (solid lines). The $c$-axis resistivity $\rho_c$ marked by the thin circle has been adopted from Ref.~\cite{intro:THz_radiation} for $T>50\ K$.  For the present calculations, $\rho_c$ has arbitrarily been extrapolated to low temperatures while imitating  a tendency to saturation observed in experiments elsewhere~\cite{materials:saturation_Takeya,materials:sub-gap_Latyshev,materials:sub-gap_we}.
The data represented by the dashed- and dash-dotted lines have been adopted from Refs.~\cite{materials:sub-gap_Latyshev} and \cite{materials:sub-gap_we}, respectively.}
\end{figure}

The $c$-axis resistivity $\rho _c$ below $T_c$ can be obtained from the sub-gap resistance $R_{sg}$ of IJJ's at different temperatures~\cite{materials:sub-gap_Latyshev,materials:sub-gap_we}. $R_{sg}$ is estimated in the limit $I,V \rightarrow 0$ and is therefore not affected by Joule heating. $R_{sg}$ shows a tendency to saturate at low temperatures possibly reflecting the presence of weakly-conducting channels short-circuiting the interlayer tunneling resistance which otherwise should ideally be infinite at $T=0$~K~\cite{materials:saturation_Takeya,materials:sub-gap_we}.

\begin{figure}
\includegraphics[width=8.5cm]{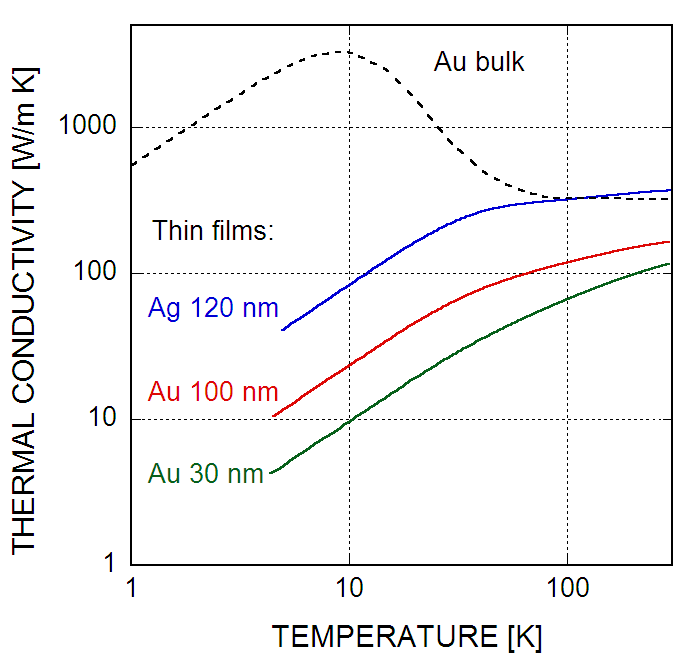}
\caption{\label{fig:gold_param}(Color online) The thermal conductivity $k(T)$ of bulk gold~\cite{materials:bulk_gold} and gold thin films of different thickness. $k(T)$ of a silver thin film is given for comparison. $k(T)$ of the thin films was calculated from the measured resistivity $\rho (T)$ by using the Wiedemann-Franz law.  Note the difference between the thermal conductivities of the thin films and the bulk gold even at room temperature~\cite{materials:gold}.}
\end{figure}

A normal-metal electrode on top of a mesa can sometimes create a delusive impression of being an effective heat-leading or heat-spreading media.  Not only the overall thermal \emph{conductance} of such a thin film is low because of its small thickness,  the thermal \emph{conductivity} of material in its bulk- and thin-film forms can be very different even at room temperature, e.g. gold~\cite{materials:gold} (see Fig.~\ref{fig:gold_param}). The mean-free path of heat carriers is limited by small grain size making thermal conductivity of polycrystalline thin films smaller than in bulk.    In the present calculations, I use the thermal conductivity of gold calculated from an experimentally measured resistivity of a 100-nm thick gold thin film by using the Wiedemann-Franz law, $k(T) = (2.44 \times 10^{-8}\ \mathrm{V^2 K^{-2}})\ T\/ \rho (T)$.

The thermal conductivity of the PMMA layer is adopted from Ref.~\cite{materials:k_PMMA}; the layer is 20~$\mu$m thick. The thermal conductivity of other possible epoxies and glues is not very different from that of PMMA~\cite{materials:C_PMMA}.

An unknown type of "silver glue" is used in the original paper \cite{intro:THz_radiation} to attach the single crystals to substrates. Note however that the thermal \emph{conductance} of such a glue layer is very likely to be smaller than that of the PMMA layer adopted in the present calculations. Indeed, admixture of either metallic and non-metallic particles to a polymer results in additional scattering of the polymer phonons at low temperature \cite{materials:k_PMMA_composite} thereby making $k$ smaller. In the case of a metallic-particle filler, the electronic part of $k$ is not high either, due to a  poor electrical connectivity between the particles. Such composite polymers have high resistivity, $\sim 0.5-1.5\ \mathrm{m\Omega \  cm}$ in most cases  \cite{materials:k_Stycast_composite} accounting for barely 10\% of the total $k$. However, viscosity of a glue with filler can be a factor of $10-100$ higher than that of the unfilled epoxy (e.g. Stycast 1265 \cite{materials:k_Stycast_composite}) which results in a much thicker layer of glue between the single crystal and substrate.

\section{\label{sec:Newton}Newton's law of cooling}

A simple self-consistent analysis of I-V's in the presence of Joule heating can be made using Newton's law of cooling:
\begin{eqnarray}
T = T_0  + \alpha I V(I,T)\label{eq:N1} \\
V(I,T) = I R(T)\label{eq:N2},
\end{eqnarray}
\noindent where $\alpha$ and $R$ are the thermal- and electrical resistances, respectively.
$R(T)$ here is assumed to be independent of $I$ i.e. I-V's are  linear (ohmic) in the absence of self-heating while $\alpha =$const with respect to both current and temperature. All the non-linearities of the overall I-V's are assumed to be due to the self-heating alone.

However, the thermal resistance $\alpha$ resulting from such an analysis is merely some average value not reflecting a real temperature distribution in the mesa. No wonder that $\alpha$ is found to be quite small for large mesas, $\alpha \approx 3.5$ K/mW ~\cite{large_heating_Newton}. This value should be compared with $\alpha \sim 50-300$ K/mW directly measured for different geometries of \emph{small} mesas and zigzag stacks of IJJ's~\cite{heating_meas_we_PRL,heating_meas_Wang_APL,heating_meas_we_SUST}.

In the case of unshunted IJJ's, the I-V's should be non-linear even without Joule heating because of tunneling between CuO planes which are d-wave superconductors. The density of states of a d-wave superconductor is linear in energy at small bias. The tunneling conductance $\Sigma(V)$ of an unshunted d-wave superconductor-insulator-superconductor (S-I-S) tunneling junction should then be parabolic in voltage. Moreover,  $\Sigma(V)$ should become even steeper on approaching the gap anomaly at high bias. Note however that for \textit{large and high} mesas the assumption of linear I-V's can be quite reasonable (see below).

The postulate that the self-heating is the \emph{only} origin of the I-V non-linearities~\cite{Zavaritsky:PRB} does not stand up well to the following qualitative and simple reasoning (see also other arguments in Ref.~\cite{Krasnov:comment_VNZ}).

The I-V curves of IJJ's consist of multiple hysteretic quasiparticle branches, each corresponding to an increasing number of IJJ's switching to the quasiparticle state with increasing current.  Suppose that the I-V's are linear without heating and $\alpha$ in Eq.~\ref{eq:N1} is big enough to account for the experimentally observed non-linearity of the first branch. Then, doubling of the dissipation power at the second branch must lead to an even stronger deviation from the linear behavior. In other words, the first and the second branches would never be similar in shape if the heating were that severe as to be the main reason for the I-V non-linearity.

Experiments show however that at least the first $10-15$ I-V branches can collapse onto a single curve if the voltage is normalized by the count number of the corresponding branch~\cite{intro:PG1}. This confirms that the branches have the same shape, the self-heating is moderate, and $\alpha$ is quite small. Of course, high current does result in a significant rise of the mesa temperature and in a cardinal change of I-V's, especially in mesas with large number of junctions.  The most typical is the "S"-shape of I-V's (back-bending) that has long time ago been recognized as resulting from both the self-heating and non-equilibrium effects~\cite{heating_Mints}.

When the temperature changes, so do the thermal conductivities of all the materials involved in the heat transfer from the junctions.  This means that $\alpha$ changes, too, and the simplicity of Eq.~\ref{eq:N1} gets broken. One should then solve the non-linear diffusion equation taking into account the temperature dependencies of the thermal conductivities and the particular geometry of a sample.  This is the purpose of the present article.

For large- and high mesas, the self-heating becomes essential already at currents much smaller than the superconducting critical current $I_c$. To simplify calculations, the I-V's can quite confidently be regarded as linear at such currents and in the absence of self-heating.  Indeed, take a mesa $S=50\times 300$~$\mu$m$^2$ large and $h=1\ \mu$m high typical for THz-emission experiments. It has about 700 IJJ's in series. The sum-gap voltage should reach the maximum value $V_{gm} \sim 40$~V assuming the typical gap voltage $v_g=2\Delta/e=60$~mV per junction. The experimental voltage range is $\sim 20$ times smaller than $V_{gm}$ due to self-heating and semiconductor-like $R(T)$. The mesa resistance $R$ at $V\rightarrow 0$ can be estimated from the $c$-axis resistivity $\rho_c \sim 600\ \Omega \mathrm{cm}$, $R_{sg} = \rho_c h/S \sim 400 \ \Omega$. Assuming that the sub-gap resistance does not change with voltage (i.e. the quasiparticle branch is linear), the current $I_{gm}$ corresponding to $V=V_{gm}$ is $\sim 40\ \mathrm{V}/400\ \Omega = 100$~mA. The range of currents used in the typical THz-emission measurements is at least two times smaller than $I_{gm}$ and corresponds to voltages $\leqslant V_{gm}/2$. In other words, it is enough to assume a linear $I(V)$-dependence up to roughly half the heating-free gap voltage to justify  the use of Newton's law of cooling. The I-V's of the small- and shallow mesas that are least affected by the self-heating indicate that this simplification is  acceptable~\cite{discussion:Zhu_small}.

Summarizing the discussion above, I will keep the reasonable assumption of linear I-V's while let the thermal resistance $\alpha$ vary with temperature in  modified Newton's law of cooling. It should be noted that Newton's law of cooling is important only for calculation of the self-consistent I-V's. The non-uniform temperature distribution can qualitatively be obtained even in the simplest case of uniform power dissipation and without any assumptions regarding the mesa resistance.

\section{\label{sec:results}Results of simulations and discussion}

The Joule heating in mesas, however trivial it may look, has not been paid due attention to until several spectroscopic studies of the pseudogap involving IJJ's appeared~\cite{intro:PG1,intro:PG2}.  These measurements required rather high bias current to reach the I-V regions beyond the superconducting gap voltage.  In the case of significant Joule heating, a normal-state resistance that decreases with temperature can result in the appearance of maxima in the $\Sigma(V)\equiv dI/dV(V)$-curves, resembling the notorious pseudogap humps~\cite{heating_meas_we_PRL,Zavaritsky:PRL}.

In general, the high-temperature superconductors have low thermal conductivity.  In Bi2212, the relatively high superconducting critical current density in the c-axis direction- ($j_c\sim 200 - 10000\ \mathrm{A\ cm^{-2}}$) and the large superconducting energy gap ($V_g\sim 50$ meV) set the scales for current and voltage. The product of the two determines the heating power density $P$ in a single IJJ: $P = 10 - 500\ \mathrm{W\ cm^{-2}}$.

To get the total power density, $P$ should be multiplied by the number of junctions $N$ in the mesa. $N$ is several hundred in the case of high ($h\sim 1\ \mu$m) mesas showing THz radiation.  An efficient removal of dissipated power becomes very important.  It is worth noting that it is not the scary-high power densities alone that determine the temperature inside the mesas. The mesa geometry, the material of the substrate, the glue, the contacts etc. are also important parameters of the problem which have been taken into account in the present calculations.

\begin{figure}
\includegraphics[width=8.5cm]{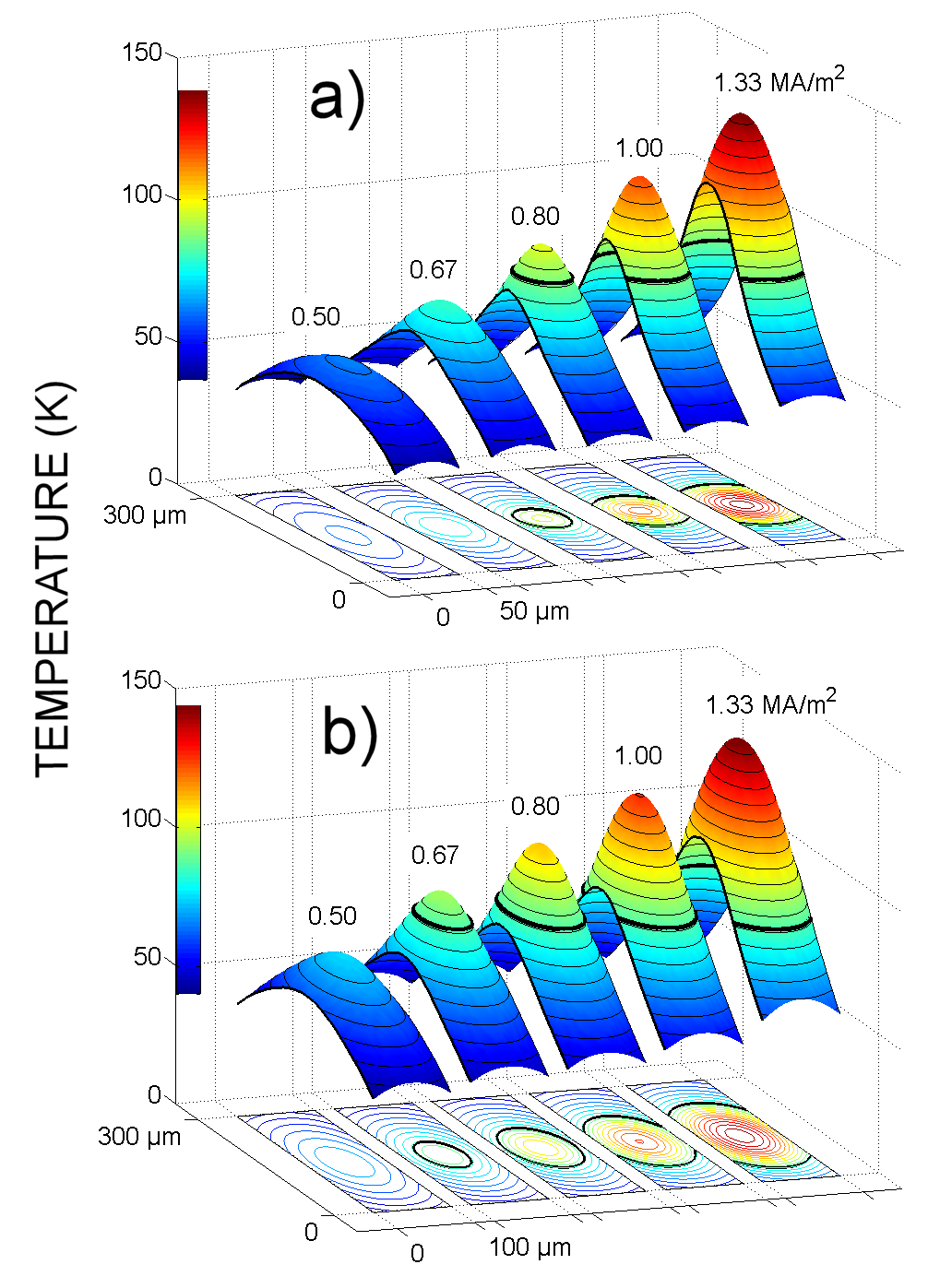}
\caption{\label{fig:Tdist_50&100}(Color online) The temperature distribution in mesas 50~$\mu$m (a) and 100~$\mu$m (b) wide at different current densities. The contour lines are drawn for every 5~K. $T=T_c$ is marked by the thick lines. Note that a significant part of the mesas can be overheated from the bath temperature $T_{0}=10$~K to temperatures above $T_c$. }
\end{figure}

Two-dimensional temperature distributions calculated for several current densities in mesas $w=50$- and 100 $\mu$m wide are shown in Fig.~\ref{fig:Tdist_50&100} a) and b), respectively.  The distributions correspond to a horizontal plane at half the mesa height. The temperature is spatially non-uniform and increases with current. For the high current densities, the temperature can even exceed $T_c$ in the middle part of the mesa. The thick contour lines indicate the boundary between the superconducting and normal-metal regions. The normal-metal part grows with current and the superconducting mesa eventually breaks up into two smaller ones connected by the common electrode and the normal-state region. The latter can provide an effective resistive shunting for the rest of the mesa which will be discussed below.

Comparison of the two panels in Fig.~\ref{fig:Tdist_50&100} also demonstrates that the wider the mesa the smaller the current density at which the normal-state region first appears. However, the maximum temperature attained at the maximum bias is not much different in all cases.

\begin{figure}
\includegraphics[width=8.5cm]{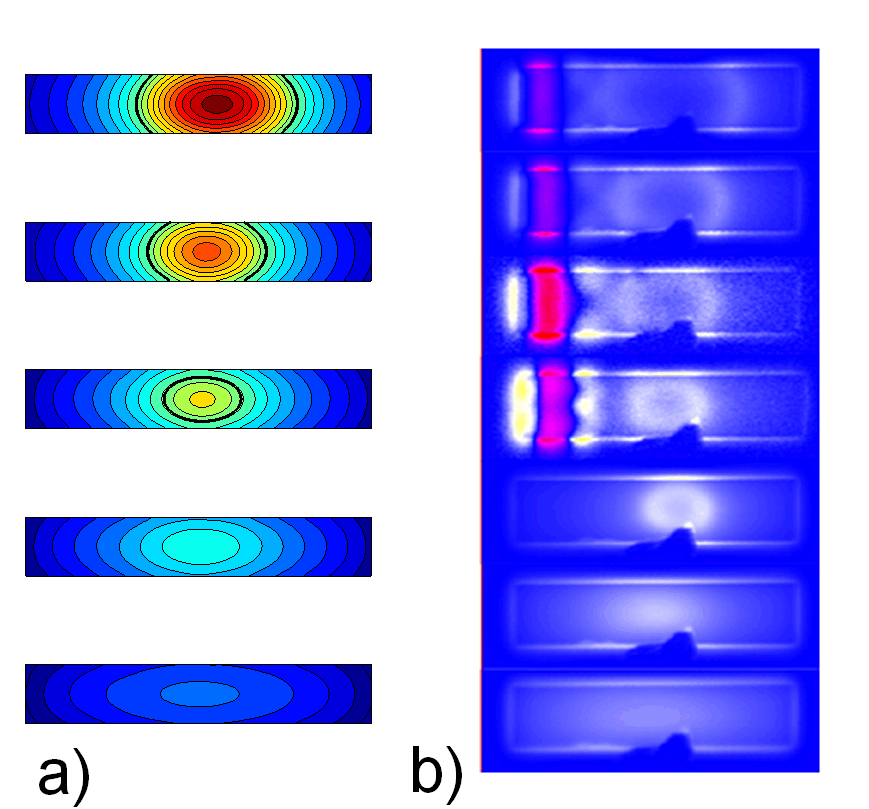}
\caption{\label{fig:Tdist_50&laser}(Color online) (a) The temperature distribution in the $50\times 300\ \mu $m$^2$ mesa at 0.5, 0.67, 0.8, 1.0, and 1.33 MA/cm$^2$, from bottom to top, respectively. The color scale and contour lines are the same as in Fig.~\ref{fig:Tdist_50&100}. (b) The experimental laser-microscopy image of the $70\times 330\ \mu$m$^2$ mesa at 0.61, 0.86, 1.05, 1.15, 1.50, 1.57, and 1.96 MA/cm$^2$ of the bias current density, from bottom to top, respectively. The false-color scale indicates the mesa-voltage variations due to the local additional heating in the laser spot which is scanned over the mesa surface. The picture has been reproduced from the Supplementary information of the paper by H.B. Wang, \emph{et al.}, Phys. Rev. Lett.\textbf{102}, 017006 (2009), Ref.~\cite{discussion:Wang_hot_spots}, $\copyright$ American Physical Society. }
\end{figure}

Fig.~\ref{fig:Tdist_50&laser} shows a side-by-side comparison of the temperature-distribution patterns obtained in the present calculations and recent laser-microscopy experiments~\cite{discussion:Wang_hot_spots}. Experimental picture shows top-electrode voltage variations due to additional heating from a small laser spot which is scanned over the mesa structure. The spot is about two microns in diameter and is estimated to raise the local temperature in the spot by a few kelvins~\cite{discussion:Wang_hot_spots}. The voltage variations are less than 0.5~mV~\cite{discussion:Wang_hot_spots} which is much smaller than the mesa voltage of $1-2$~V.

The laser-microscopy patterns reveal distinct rings with their diameter increasing with bias current.  The rings resemble thick contour lines of the simulations that correspond to $T=T_c$. It is fully reasonable to expect large resistance variations at local places where temperature is close to $T_c$, giving rise to relatively larger signals than from other parts of the mesa in the laser-microscopy experiments.

At some high bias current the laser microscope reveals the presence of some kind of standing waves giving very strong response in the experiment (see the pink spots in Fig.~\ref{fig:Tdist_50&laser}b. The origin of the spots is not totally clear and is now being debated. This is however beyond the scope of the present work and will not be discussed here~\cite{discussion:comment_precision}.

\begin{figure}
\includegraphics[width=8.5cm]{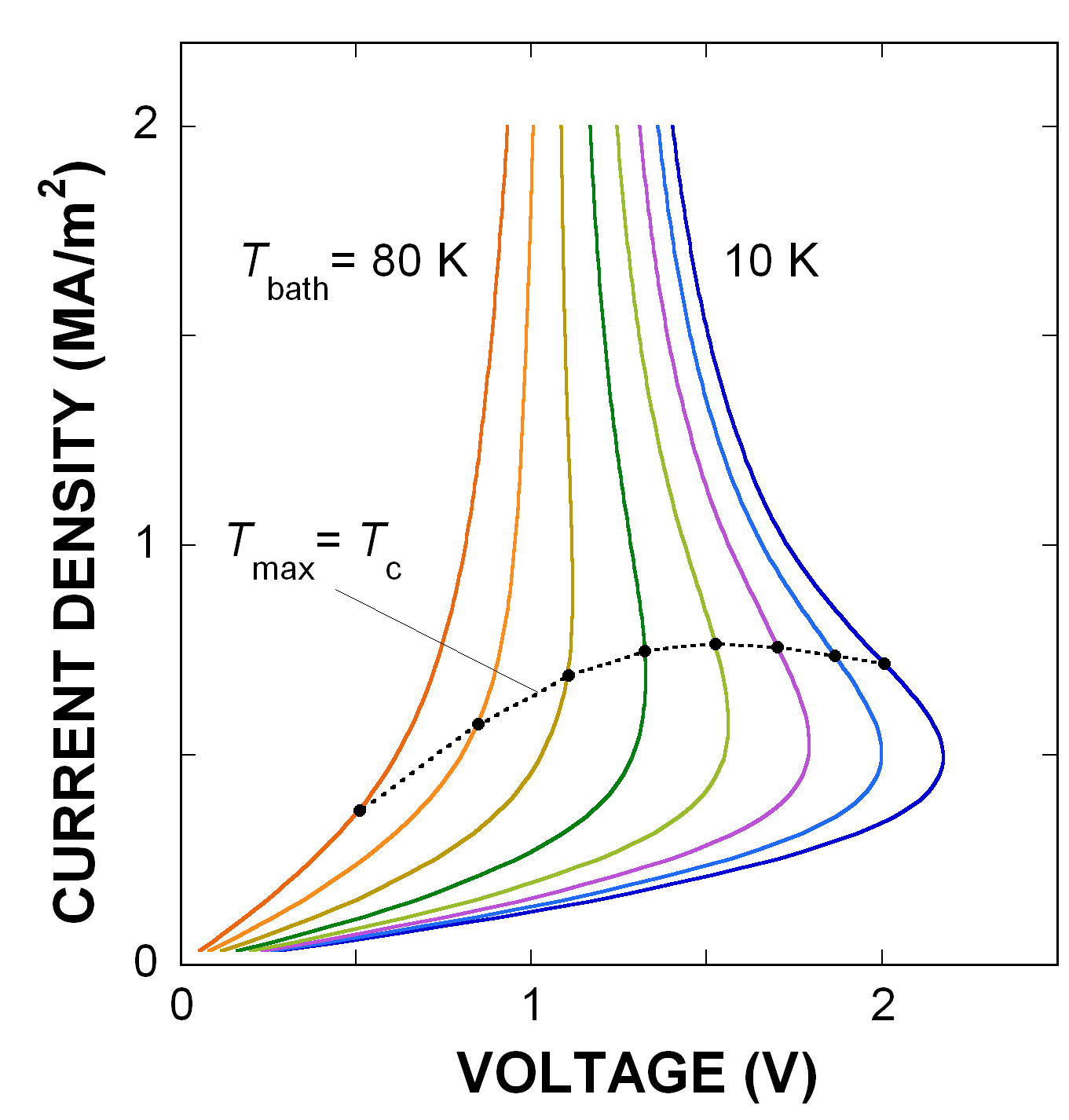}
\caption{\label{fig:V(I)_50}(Color online) The current-voltage characteristics of the 50-$\mu$m-wide mesa at different bath temperatures from 10 to 80 K with 10-K intervals (solid lines). The dashed line indicates a boundary at which the maximum of the temperature distribution in the mesa reaches $T_c$.}
\end{figure}

Fig.~\ref{fig:V(I)_50} shows I-V's for a mesa 50~$\mu$m wide at different bath temperatures from 10 to 80 K with  10-K intervals. The I-V's have been calculated self-consistently, assuming the three-dimensional heat- and electrical-current distributions.   The dashed line indicates the boundary at which the maximum temperature $T_{max}$ of the temperature distribution reaches $T_c$. The curves  reproduce experimental I-V's qualitatively well~\cite{intro:THz_radiation}. It is also seen that the back-bending decreases with $T_{0}$, also in accordance with experiments.

\begin{figure}
\includegraphics[width=8.5cm]{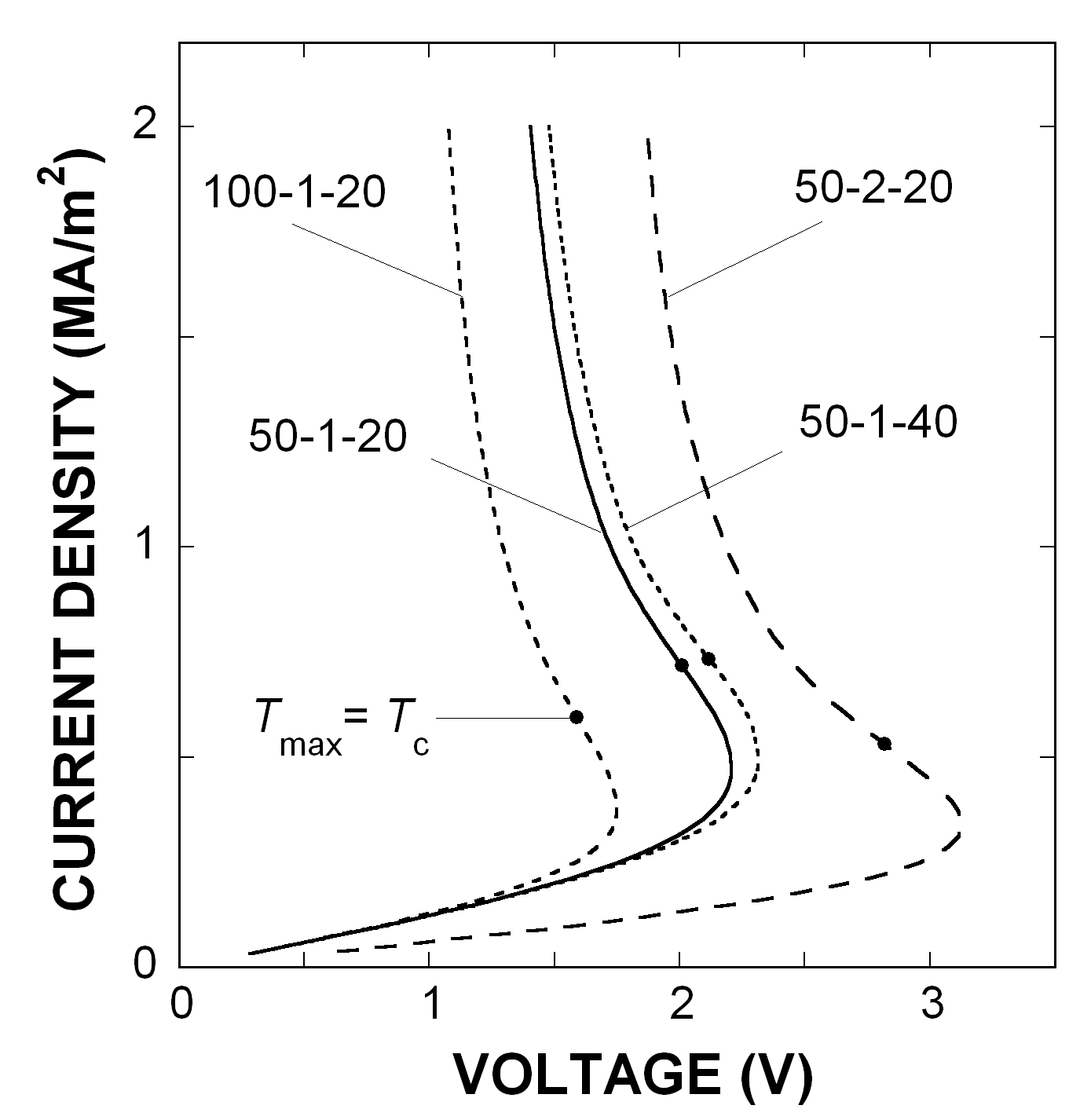}
\caption{\label{fig:Tdist_sizes} The current-voltage characteristics of the 50-$\mu$m-wide mesa at $T_{0}=10$ K. The triple numbers indicate the mesa width, the mesa height, and the thickness of the underlying single crystal (in $\mu$m). The dots mark the I-V points where $T_{max}=T_c$.}
\end{figure}

Fig.~\ref{fig:Tdist_sizes} shows calculated I-V's for different mesas at $T_{0}=10$ K. The triple numbers along each line indicate the mesa width, height, and the thickness of the underlying single crystal. The solid dots indicate the I-V points where $T_{max}=T_c$.

It is the mesa sizes that largely govern the maximum temperature whereas the thickness of the underlying single crystal is of minor importance. Indeed,  compare curve "50-1-20" versus "50-1-40", and "50-1-20" versus "50-2-20" or "100-1-20".  Qualitatively,  the heat flow below the mesa into a half-space of the single crystal can be modeled as having cylindrical or even spherical symmetry, depending on the mesa length-to-width ratio. The thermal conductivity of Bi2212 is anisotropic, with $k_{ab}/k_c\sim 10$. To roughly map the heat-transfer problem to an equivalent isotropic media, the effective thickness of the single crystal should be proportionally increased making it larger than the width of the mesa ($d'\sim k_{ab}/k_c\ d = 200 \ \mathrm{\mu m}> w$, see Fig.~\ref{fig:geometry}). In the cylindrical geometry, the largest temperature gradient is close to the heat source. Making the heat-transfer media around the source much thicker than the source itself has therefore little effect on its temperature.

From Figs.~\ref{fig:V(I)_50} and \ref{fig:Tdist_sizes} it is seen that $T_{max}=T_c$ at a bias current slightly above the current $I_g$ corresponding to the maximum voltage $V_g$ of the I-V's with negative differential conductance. This is in agreement with the direct measurements of the mesa temperature~\cite{heating_meas_we_PRL}.

The normal-state region extends by about 5 $\mu$m deep into the pedestal of the mesa at high current (see Fig.~\ref{fig:shunting}).  This indicates that the effective cooling of the mesa can only be achieved if the pedestal is made thinner than that size.

Recently, THz radiation with a relatively high output power has been observed at currents appreciably higher than  $I_g$~\cite{intro:THz_strong} as compared with the original experiments where a much weaker radiation was measured at $I < I_g$~\cite{intro:THz_radiation}.   Very stable  high-power radiation that extends to four harmonics (corresponding to 2.5 THz), all having a relatively narrow line width, have been observed in the experiments.

\section{Synchronization of radiation}

The fact that the mesa temperature can exceed $T_c$ at $I > I_g$ where the intense THz radiation has been observed makes it reasonable to suggest that the heating plays an active role in the synchronization of IJJ's in large mesas.

\begin{figure}
\includegraphics[width=8.5cm]{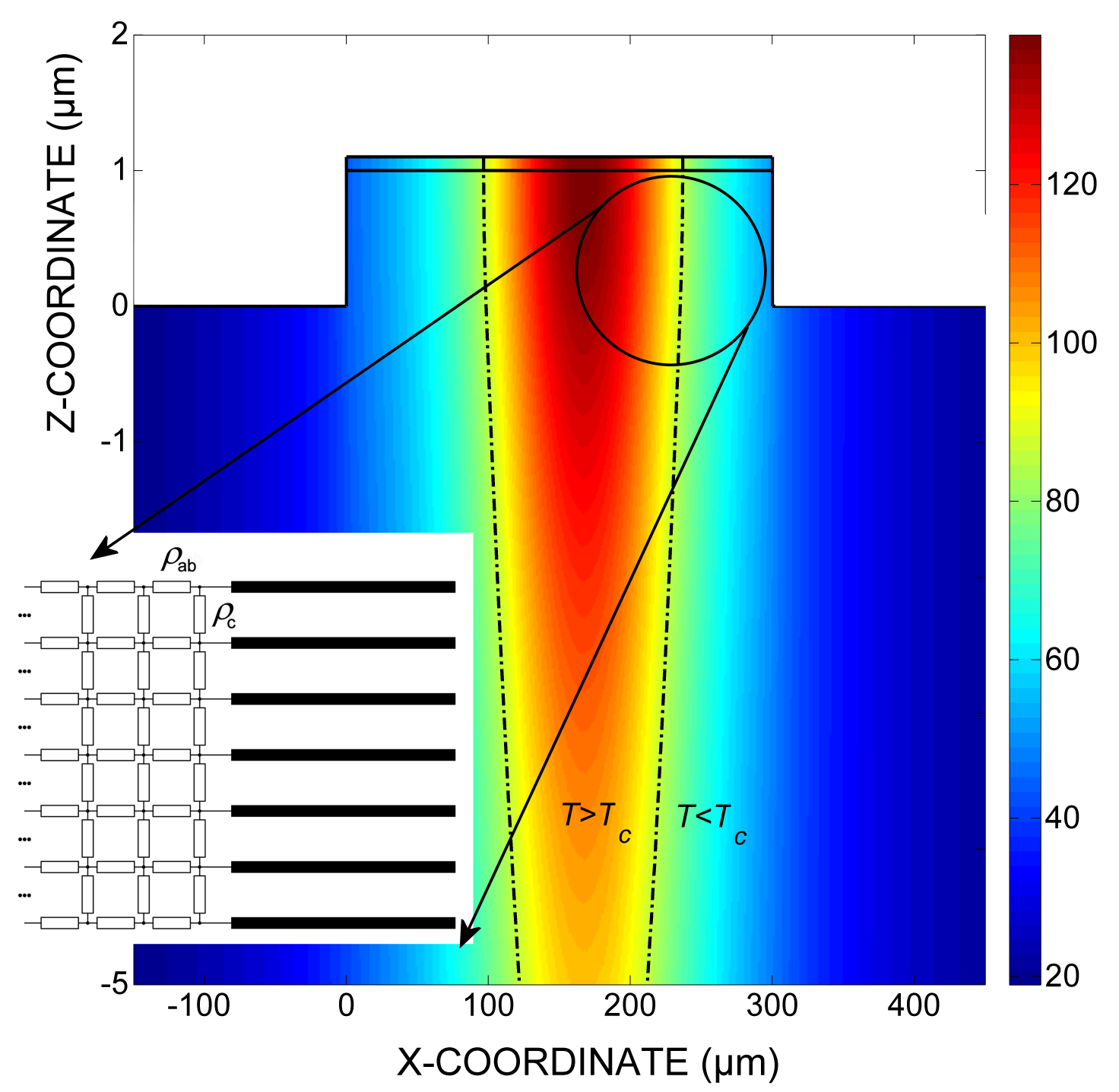}
\caption{\label{fig:shunting}(Color online) The temperature distribution across the vertical middle section of the mesa and the underlying single crystal at $j=1.33$~MA/cm$^2$. The dash-dotted line marks the isotherm $T=T_c$.  The inset schematically shows a simplified picture of the resistively shunted section of the mesa. The equivalent shunting circuitry is an infinite matrix of resistors representing the in- and out-of-plane resistivities.}
\end{figure}

Indeed, the normal-state part of the mesa at $I > I_g$ can effectively create a shunting resistor for the rest of the mesa which is still in the superconducting state.  Even in the superconducting parts, the temperature is non-uniform, varying from close to the bath temperature up to $T_c$ over some 100 $\mu$m (see Fig.~\ref{fig:shunting}). Both the shunting resistance and an uneven critical current density resulting from such a non-uniform temperature distribution can presumably help synchronization of the Josephson junctions in the mesa.

It is known that a load impedance common to all the Josephson junctions in a series array can stimulate their synchronization~\cite{shunted_arrays_J}. Resistive, capacitive, and inductive  loads and some of their combinations have been considered in Ref.~\cite{shunted_arrays_J}. In general, the overdamped Josephson junctions (with $\beta_c \equiv 2\pi I_c R_n^2 C/\Phi_0 \sim 1$) are easier to synchronize by the common load. It is not straightforward to assess the equivalent $\beta_c$ for IJJ's with the critical current density and the normal-state resistance being strong functions of coordinates.

To estimate the effective shunting resistance resulting from the normal-state parts of the mesa, I divide the average voltage drop across the mesa height by the total current flowing through the normal-state area, $I_n$. The voltage does not change much along the mesa upper surface and can be assumed to be constant.  I integrate the spatially varying current density over the normal-state area to find $I_n$.  The effective shunting resistance appears to quickly decrease with bias current down to $100-200\ \Omega$ which is equivalent to $0.15-0.3\ \Omega$ per junction. This value is of the order of the normal-state resistance estimated for a $50\times 100 \ \mu \mathrm{m}^2$-large IJJ at $T\lesssim T_c$.

The suggestive active role of the heating-induced resistive shunting in synchronization of the radiation can possibly be proven experimentally. The intrinsic resistive shunting can be created even for relatively small currents $I<I_g$ by intentionally destroying superconductivity in some parts of the mesa by, say, a local Si-ion implantation~\cite{discussion:Si-implantation}. The radiation is then expected to occur for a wider range of bias currents and temperatures than in Refs.~\cite{intro:THz_radiation,intro:THz_strong}.

It is interesting to compare the intrinsic shunting described above with traditional techniques. A resistive shunting of IJJ's by an external resistor would imply deposition of a thin metal film connecting the top and bottom of the mesa. Effective resistance of the shunt is then decided by the relatively high contact resistance $R_{\mathrm cont}$ between the metal and superconductor. Firstly, a sufficiently low $R_{\mathrm cont}$ can only be obtained when the metal thin film is deposited on the freshly cleaved surface of Bi2212. The mesa vertical sides, as well as the etched down parts of the single crystal are likely to be significantly degraded by the lithography processing and etching thereby not allowing for low contact resistances. Secondly, the butt-end contact area of  individual IJJ electrodes, away from the mesa top or bottom, is very small, of the order of the electrode thickness times the mesa width, $\sim 3{\AA} \times 50\ \mu \mathrm{m}=1.5\times 10^{-14}$ m$^2$. Even if taking the most optimistic contact resistivity that has been observed in experiments, $8\times10^{-5}\ \mathrm{\Omega \ cm^2}$ \cite{contact_R}, the $R_{\mathrm cont}$ can hardly be smaller than about 50 k$\Omega$. This possibly explains an earlier unsuccessful attempt to synchronize IJJ's in Bi2212 mesas by the external shunting resistance~\cite{shunted_IJJ's}.

Even well below $I_g$, the temperature is distributed non-uniformly (see Fig.~\ref{fig:Tdist_50&100}). The spatial variation of $j_c$ can be a reason for synchronization of radiation, as has been suggested in Ref.~\cite{intro:THz_model1}. However, the radiation should then correspond to only even frequency modes because of a nearly symmetric temperature- and $j_c$ variations for small bias currents. To account for the frequencies observed in the experiments~\cite{intro:THz_radiation}, a mesa with twice the width should have been used~\cite{intro:THz_model1}.

Nonetheless, at $I\gtrsim I_g$, when the mesa is subdivided into three parts, the symmetry along the mesa length gets broken. Each of the superconducting parts now has $j_c$ varying from zero at the side facing the overheated region and up to an almost nominal $j_c$ at the opposite side (see Fig.~\ref{fig:shunting}).  A difference with Ref.~\cite{intro:THz_model1} is in that $j_c$ here varies along the \emph{length} of the mesa  and  it is now two effective mesas having the common load and top electrode that should be considered. Although it is plausible to expect the synchronization even for this geometry, detailed analysis of this situation is beyond the scope of this article.

\section{Summary}
Self-heating in large-area high mesas that are used in experiments on powerful THz emission from IJJ's is analyzed by numerically solving the diffusion equation.  The current-voltage characteristics are calculated self-consistently showing regions of negative differential conductance. Particular mesa geometry and the temperature dependencies of all the electrical- and thermal materials parameters are included in the calculations. The calculations reveal that  the mesa temperature is not only significantly higher than the bath temperature, it is strongly non-uniform along the mesa length and width.  The middle part of the mesa can be overheated even above $T_c$ at high bias that is consistent with the laser-microscopy experiments. The normal-state regions make up an effective resistive shunting for the rest of the mesa. The non-uniform temperature distribution and/or the resistive shunting  can possibly be important for synchronization of emission from the junctions in the mesa.

\begin{acknowledgments}
I acknowledge support from the Swedish Research Council through the Linnaeus centrum "Engineering quantum systems" and wish to thank J. F. Schneiderman for reading the manuscript.
\end{acknowledgments}



\begin{references}
\bibitem{discussion:comment_precision} It is in principle possible to include the local heating by laser in the simulations. However, the precision is not high enough to resolve some 100~$\mu$V-variations in voltage that is of the order of several volts. Very small mesh size would probably be needed to reproduce experimental results. This would than require a giant computer memory and unreasonably long time to perform such calculations.
\end{references}
\end{document}